\documentclass[letterpaper]{acm_proc_article-sp}
\usepackage[
  pass,
]{geometry}
\usepackage{changepage}
\usepackage{color}
\usepackage{graphicx}
\usepackage{url}
\usepackage{adjustbox}
\usepackage{booktabs}
\usepackage{array}
\usepackage{balance} 
\usepackage{paralist} 
\usepackage[caption=false]{subfig}
\newcolumntype{M}{>{\centering\arraybackslash}m{0.6in}}
\usepackage{multirow}
\usepackage{rotating}
\permission{
}
 
\begin{document}

\title{Is Saki \#delicious? The Food Perception Gap on Instagram and Its Relation to Health
\titlenote{This is a pre-print of our paper accepted to appear in the Proceedings of 2017 International World Wide Web Conference (WWW'17).}
}

%
%
%
%
%

\numberofauthors{5} 
%
\author{
%
%
\alignauthor
Ferda Ofli\\
       \affaddr{Qatar Computing Research Institute, HBKU}\\
       \affaddr{Doha, Qatar}\\
       \email{fofli@hbku.edu.qa}
\alignauthor
Yusuf Aytar\\
       \affaddr{CSAIL}\\
       \affaddr{MIT}\\
       \affaddr{Cambridge, USA}\\
       \email{yusuf@csail.mit.edu}
\alignauthor 
Ingmar Weber\\
       \affaddr{Qatar Computing Research Institute, HBKU}\\
       \affaddr{Doha, Qatar}\\
       \email{iweber@hbku.edu.qa}
\and  
\alignauthor
Raggi al Hammouri\\
       \affaddr{Qatar Computing Research Institute, HBKU}\\
       \affaddr{Doha, Qatar}\\
       \email{ralhammouri@hbku.edu.qa}
\alignauthor 
Antonio Torralba\\
       \affaddr{CSAIL}\\
       \affaddr{MIT}\\
       \affaddr{Cambridge, USA}\\
       \email{torralba@mit.edu}
}

\maketitle
\begin{abstract}
Food is an integral part of our life and what and how much we eat crucially affects our health. Our food choices largely depend on how we perceive certain characteristics of food, such as whether it is healthy, delicious or if it qualifies as a salad. But these perceptions differ from person to person and one person's ``single lettuce leaf'' might be another person's ``side salad''. Studying how food is perceived in relation to what it actually is typically involves a laboratory setup. Here we propose to use recent advances in image recognition to tackle this problem. Concretely, we use data for 1.9 million images from Instagram from the US to look at systematic differences in how a machine would \emph{objectively} label an image compared to how a human \emph{subjectively} does. We show that this difference, which we call the ``perception gap'', relates to a number of health outcomes observed at the county level. To the best of our knowledge, this is the first time that image recognition is being used to study the ``misalignment'' of how people describe food images vs.\ what they actually depict.
\end{abstract}

\keywords{Instagram; food; public health; computer vision}

\section{Introduction}\label{sec:introduction}
Food is a crucial part of our life and even our identity. Long after moving to a foreign country and after adopting that country's language, migrants often hold on to their ethnic food for many years \cite{fischler88food}. Food is also a crucial element in effecting weight gain and loss, with important implications on obesity and diabetes and other lifestyle diseases. Some researchers go as far as claiming that ``you cannot outrun a bad diet''~\cite{Malhotra22042015}.

One important aspect governing our food choices and how much we consume is how we perceive the food. What do we perceive to be healthy? Or delicious? What qualifies as a ``salad''? Food perception is typically studied in labs, often using MRIs and other machinery to measure the perception at the level of brain activity \cite{killgore2005body,rosenbaum2008leptin,medic2016presence}. Though such carefully controlled settings are often required to remove confounding variables, these settings also impose limitations related to (i) the artificial setting the subject is exposed to, and (ii) the cost and lack of scalability of the analysis.

There are, however, externally visible signals of food perception ``in the wild'' that can be collected at scale and at little cost: data on how people label their food images on social media. What images get labeled as \#salad? Which ones get the label \#healthy?

Though useful, these human-provided labels are difficult to disentangle from the actual food they describe: if someone labels something as \#salad is this because (i) it really is a salad, or (ii) the user believes that a single lettuce leaf next to a big steak and fries qualifies as a salad.


\begin{figure}[h!]
  \setlength{\tabcolsep}{0pt}
  \begin{tabular}{ m{\columnwidth} }
    \begin{minipage}{\columnwidth}
      \includegraphics[width=\columnwidth]{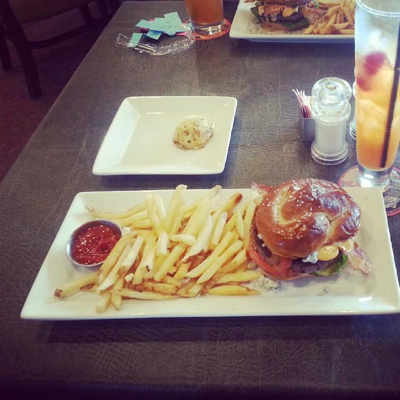}
    \end{minipage}
    \\
      \begin{itemize}[]
        \item \underline{User}: \#foodie, \#hungry, \#yummy, \#burger
        \item \underline{Machine}: \#burger, \#chicken, \#fries, \#chips, \#ketchup, \#milkshake
      \end{itemize}
  \end{tabular}
  \caption{A comparison of user-provided tags vs.\ machine-generated tags. In this example, the user uses only \#burger to describe what they are eating, potentially not perceiving the fries as worth mentioning, though they are providing subjective judgement in the form of \#yummy. However, machine-generated tags provide more detailed factual information about the food plate and scene including \#fries, \#ketchup, and \#milkshake.}
  \label{fig:gap_ex1}
  \end{figure}
  
\begin{figure}[h!]
  \setlength{\tabcolsep}{0pt}
  \begin{tabular}{ m{\columnwidth} }
    \begin{minipage}{\columnwidth}
      \includegraphics[width=\columnwidth]{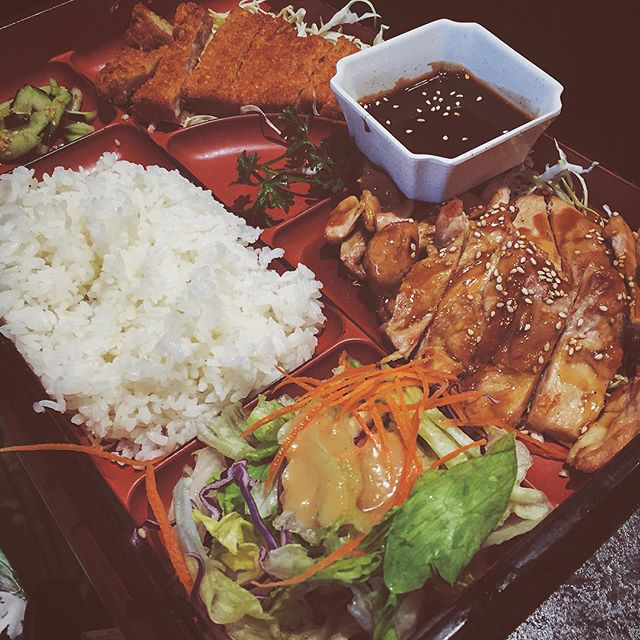}
    \end{minipage}
    \\
      \begin{itemize}[]
        \item \underline{User}: \#pork, \#foodie, \#japanese, \#chicken, \#katsu, \#box, \#salad, \#restaurant, \#bento, \#rice, \#teriyaki
        \item \underline{Machine}:  \#teriyakichicken, \#stickyrice, \#chickenkatsu, \#whiterice, \#teriyaki, \#peanutsauce, \#ricebowl, \#spicychicken, \#fishsauce, \#bbqpork, \#shrimptempura, \#ahituna, \#friedshrimp, \#papayasalad, \#roastpork, \#seaweedsalad, \#chickenandrice%
      \end{itemize}
  \end{tabular}
  \vspace{-.2in}
  \caption{Another comparison of user-provided tags vs.\ machine-generated tags. Machine-generated tags provide more detail about the food plate such as the type of rice, i.e., \#stickyrice and \#whiterice, and the dressing on the food item, i.e., \#peanutsauce.}
  \label{fig:gap_ex2}
  \end{figure}
  

We propose to use image recognition to study the ``perception gap'', i.e., the difference between what a food image objectively depicts (as determined by machine annotations) and how a human describes the food images (as determined from the human annotations). Figures~\ref{fig:gap_ex1} and \ref{fig:gap_ex2} show examples from our dataset.

We find that there are systematic patterns of how this gap is related to variation in health statistics. For example, counties where users are, compared to a machine, more likely to use the hashtag \#heineken are counties with a higher Food Environment Index. In this particular example, a plausible hypothesis is that users who are specific about how they choose - and label - their beer are less likely to drink beer for the sake of alcohol and more likely to drink it for its taste.


We then extend our analysis to also include subjective labels applied by humans. Here we find that, e.g., labeling an image that depicts saki (as determined by the machine) as \#delicious (by the human) is indicative of lower obesity rates. This again illustrates that not only the perception of alcohol, as a fun drug to get high vs.\ as part of a refined dining experience, can be related to health outcomes, but also that such perception differences can be picked up automatically by using image recognition.

The rest of the paper is structured as follows. In the next section we review work related to (i) the perception of food and its relationship to health, (ii) using social media for public health surveillance, and (iii) image recognition and automated food detection. Section~\ref{sec:data} describes the collection and preprocessing of our Instagram datasets, including both our large dataset of $1.9M$ images used to analyze food perception gap and its relation to health, as well as even larger dataset of $\sim3.7M$ images used to train and compare our food-specific image tagging models against the Food-101 benchmark. Section~\ref{sec:computervision} outlines the architecture we used for training our food recognition system and shows that it outperforms all reported results on the reference benchmark. Our main contribution lies in Section~\ref{sec:methods} where we describe how we compute and use the ``perception gap''. Our quantitative results, in the form of indicative gap examples, are presented in Section~\ref{sec:results}. In Section~\ref{sec:discussion} we discuss limitations, extensions and implications of our work, before concluding the paper.

\section{Related Work}\label{sec:related}
Our research relates to previous work from a wide range of areas. In the following we discuss work related to (i) food perception and its relationship to health, (ii) using social media for public health tracking, and (iii) image recognition and automated food detection.




\textbf{Food perception and its relationship to health.} Due to the global obesity epidemic, a growing number of researchers have studied how our perception of food, both before and during its consumption, relates to our food choices and the amount of food intake. Here we review a small sample of such studies.

Killgore and Yurgelun-Todd~\cite{killgore2005body} showed a link between differences in orbitofrontal brain activity and (i) viewing high-calorie or low-calorie foods, and (ii) the body mass index of the person viewing the image. This suggests a relationship between weight status and responsiveness of the orbitofrontal cortex to rewarding food images.

Rosenbaum et al.~\cite{rosenbaum2008leptin} showed that, after undergoing substantial weight loss, obese subjects demonstrated changes in brain activity elicited by food-related visual cues. Many of these changes in brain areas known to be involved in the regulatory, emotional, and cognitive control of food intake were reversed by leptin injection.

Medic et al.~\cite{medic2016presence} examined the relationship between goal-directed valuations of food images by both lean and overweight people in an MRI scanner and food consumption at a subsequent all-you-can-eat buffet. They observed that both lean and overweight participants showed similar patterns of value-based neural responses to health and taste attributes of foods. This suggests that a shift in obesity may lie in how the presence of food overcomes prior value-based decision-making.

Whereas the three studies discussed above studied the perception at the level of brain activity, our own work only looks at data from perception reported in the form of hashtags. This, indirectly, relates to a review by Sorensen et al.~\cite{sorensenetal03ijo} of studies on the link between the (self-declared) palatability, i.e., the positive sensory perception of foods, and the food intake. All of their reviewed studies showed that increased palatability leads to increased intake. In Section~\ref{sec:methods:subjectivegap}, we study a similar aspect by looking at regional differences in what is tagged as \#delicious and how this relates to obesity rates and other health outcomes.

More directly related to the \emph{visual} perception of food is work by Delwiche who described how visual cues lead to expections through learned associations and how these influence the assessment of the taste and flavor of food~\cite{delwiche12pb}. For example, when taste- and odor-less food coloring is used the perceived taste of the food changes and white wine colored as red wine would begin to taste like a red wine.

McCrickerd and Forde~\cite{OBR:OBR12340} focused on the role of both visual and odor cues in identifying food and guiding food choice. In particular, they described how the size of a plate or a bowl or the amount of food served effect the food intake. Generally, larger plates lead to more food being consumed.

Closer to the realm of social media is the concept of ``food porn''. Spence et al.~\cite{spenceetal15bc} discussed the danger that our growing exposure to such beautifully presented food images has detrimental consequences in particular on a hungry brain. They introduce the notion of ``visual hunger'', i.e., the desire to view beautiful images of food.

Petit gave a more positive view regarding the potential of food porn and social media images and discusses their use in carefully crafted ``multisensory mental simulation''~\cite{petitcan}. He argued that by engineering an appropriate pre-eating experience involving images and other sensory input food intake can be reduced and healthy food choices can be encouraged.

Note that our current analysis does not look at the \emph{presentation} aspect of food images. It would, however, be interesting and technically feasible to use computer vision to extract information on how the food is presented and then attempt to link this back to health statistics.

\textbf{Social media data for public health analysis.} Recent studies have shown that large scale, real time, non-intrusive monitoring can be done using social media to get aggregate statistics about the health and well being of a population~\cite{dredze12is,sarker15jbi,kostkova13www}. Twitter in particular has been widely used in studies on public health~\cite{pauldredze11icwsm,prier2011identifying,parkeretal13asonam,Kostkova15TwitterSocioscope}, due to its vast amount of data and the ease of availability of data. 

Connecting the previous discussion on the perception of food and food images to public health analysis via social media is work by Mejova et al.~\cite{mejovaetal16icwsm}. They study data from 10 million images with the hashtag \#foodporn and find that, globally, sugary foods such as chocolate or cake are most commonly labeled this way. However, they also report a strong relationship (r=0.51) between the GDP per capita and the \#foodporn-healthiness assocation.

In the work most similar to ours, Garimella et al.~\cite{garimellaetal16chi} use image annotations obtained by Imagga\footnote{\url{http://imagga.com/auto-tagging-demo}} to explore the value of machine tags for modeling public health variation. They find that, generally, human annotations provide better signals. They do, however, report encouraging results for modeling alcohol abuse using machine annotations. Furthermore, due to their reliance on a third party system, they could only obtain annotations for a total of 200k images. Whereas our work focuses on the \emph{differences} in how machines and humans annotate the same images, their main focus is on building models for public health monitoring.

Previously, Culotta~\cite{culotta14chi} and Abbar et al.~\cite{abbaretal15chi}
	used Twitter in conjunction with psychometric lexicons such as LIWC and PERMA to predict county-level health statistics such as obesity, teen pregnancy and diabetes. Their overall approach of building regression models for regional variations in health statistics is similar to ours. 
Paul et al.~\cite{pauldredze14pone} make use of Twitter data to identify health related topics and use these to characterize the discussion of health online. 
Mejova et al.~\cite{mejovaetal15dh} use Foursquare and Instagram images to study food consumption patterns in the US, and find a correlation between obesity and fast food restaurants.

%
%
%
%

Abdullah et al.~\cite{abdullah2015collective} use smile recognition from images posted on social media to study and quantify the overall societal happiness. Andalibi et al.~\cite{andalibi2015depression} study depression related images on Instagram and ``establish[ed] the importance of visual imagery as a vehicle for expressing aspects of depression". Though these papers do not explicitly try to model public health statistics, they illustrate the value of image recognition techniques in the health domain. In the following we review computer vision work in more depth.

\textbf{Image recognition and automated food detection.} Although images and other rich multimedia form a major chunk of content being shared in social media, almost all the methods above rely on textual content.  
Automatic image annotation has greatly improved over the last couple of years, owing to the recent development in deep learning~\cite{krizhevsky2012imagenet,simonyan2014very,he2015deep}. 
Robust object recognition~\cite{wu2015deep,chatfield2014return} and image captioning~\cite{karpathy2014deep} have become possible because of these new developments. For example, Karpathy et al.~\cite{karpathy2014deep} use deep learning to produce descriptions of images, which compete with (and sometimes beat) human generated labels. 
A few studies already make use of these advances to  identify~\cite{kawanoyanai15mta,Myers:2015:ITA:2919332.2919978,wang2015recipe,liu2016deepfood} and study~\cite{sudoetal14ubicomp} food consumption from pictures. For instance on the Food-101 dataset~\cite{bossard2014food}, one of the major benchmarks on food recognition, the classification accuracy improved from $50.76\%$~\cite{bossard2014food} to $77.4\%$~\cite{liu2016deepfood} and $79\%$~\cite{Myers:2015:ITA:2919332.2919978} in recent years with the help of deep convolutional networks. 

Building upon Food-101 dataset, Myers et al.~\cite{Myers:2015:ITA:2919332.2919978} explore in-depth food understanding, including food segmentation and food volume estimation in plates, as well as predicting the calories from food images collected from restaurants. Unfortunately, the segmentation and depth image annotations used in their work are not publicly shared and cannot be used as a benchmark.


In addition to the Food-101 dataset, which has 101 classes and $101K$ images, there are various other publicly available smaller datasets, such as: PFID~\cite{chen2009pfid}, which has 61 classes (of fast food) and 1,098 images; UNICT-FD889~\cite{farinella2014benchmark}, which has 889 classes and 3,853 images; and UECFOOD-100~\cite{matsuda2012recognition}, which has 100 classes, and 9,060 images; later this dataset is expanded to 256 food categories~\cite{kawano14c}. Unfortunately, the performance for image recognition in general and food recognition in particular is highly correlated with the size of the datasets, especially while training deep convolutional models. In our work, we train deep convolutional networks with noisy but very large scale datasets collected from Instagram images. 

Rich et al.~\cite{Rich:2016:TBA:2896338.2897734} learn food classifiers by training an SVM over a set of extracted features from $\sim800K$  images collected from Instagram. Our auto-tagger is mainly different from theirs in three major components, a) we only use food-related hashtags cleaned through a crowd-sourced process, b) we train a state-of-the-art deep network on food classification rather than operating on extracted features, c) we build upon a much larger image dataset($\sim 3.7M$ images in $1,170$ categories).









\section{Data Collection}\label{sec:data}

\begin{figure}[ht]
  \centering
  \includegraphics[width=\columnwidth]{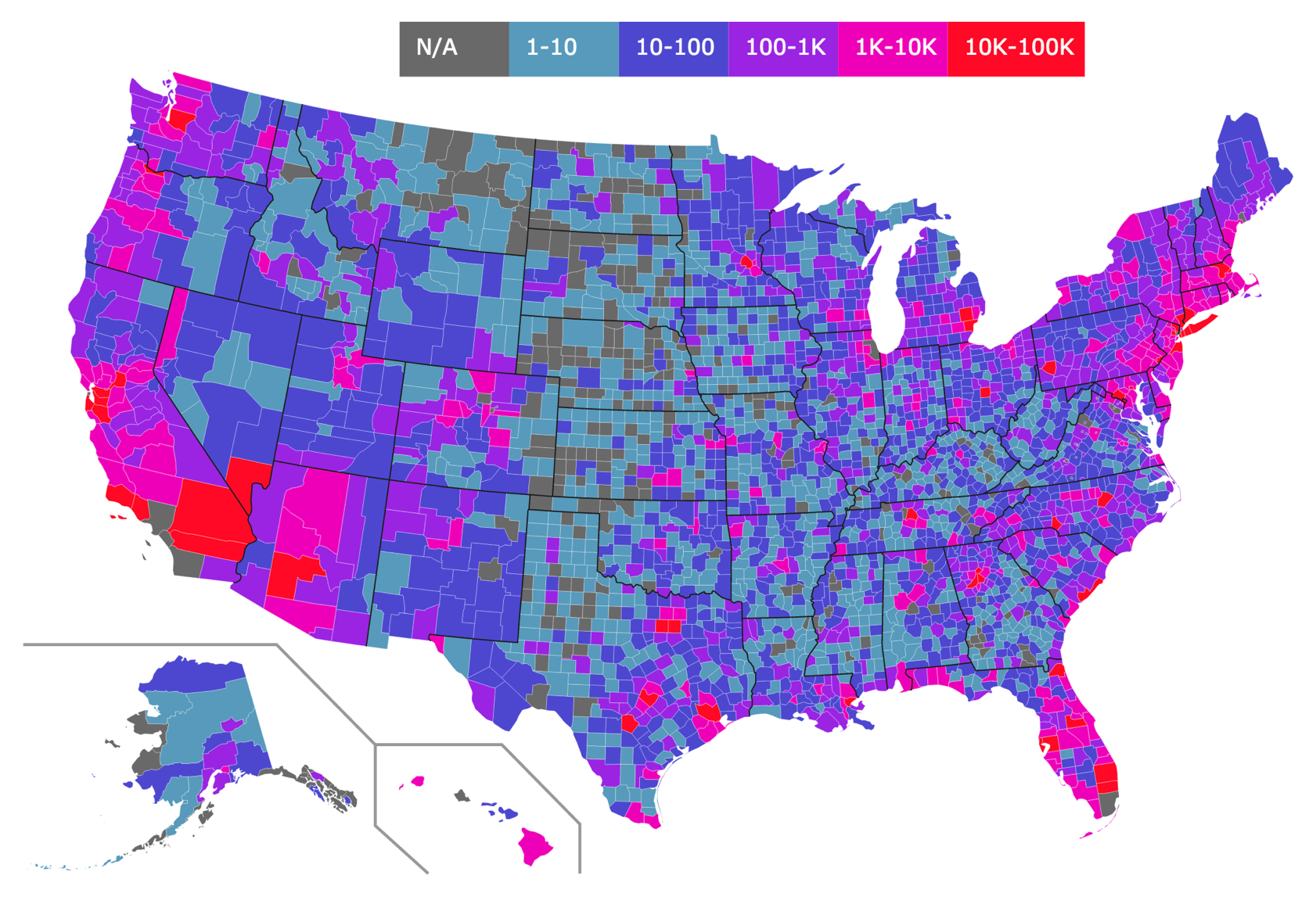}\\
  \caption{Distribution of collected $\sim4M$ images across US counties.}
  \label{fig:us_image_counties}
\end{figure}



\textbf{Instagram data collection.} In early 2016 we collected worldwide data from Instagram covering timestamps between November 2010 and May 2016 for the following hashtags: \#food, \#foodporn, \#foodie, \#breakfast, \#lunch, \#dinner. This led to meta information for a total of $\sim72M$ distinct images, $\sim26M$ of which  have associated locations, and $\sim4M$ of them are successfully assigned to one of the US counties. 
Assignments are achieved by matching the longitude and latitude information of images with the polygons for each county in the US. Computations are performed in python using the Shapely\footnote{\url{https://github.com/Toblerity/Shapely}}
package for geometric processing. The polygons are obtained from the website of the US Census Bureau\footnote{\url{https://www.census.gov/geo/maps-data/data/kml/kml_counties.html}}.
The distribution of images over the US counties are visualized in Figure~\ref{fig:us_image_counties}.


\textbf{Clean food-related hashtags.} From the collected data, the top 10,000 hashtags are extracted from the Instagram posts in the US. Since we are mainly concerned with the food perception, we manually classified these hashtags into food-related categories with the help of crowd sourcing through Amazon Mechanical Turk\footnote{\url{https://www.mturk.com/mturk}} services. Each hashtag is seen by five unique workers, and classified into the following categories: drinks, part-of-a-dish, name-of-a-dish, other-food-related, and non-food-related.
The hashtags belonging to the categories of \emph{drinks}, \emph{part-of-a-dish}, and \emph{name-of-a-dish} are joined together in order to compose our dictionary of interest for food perception analysis. This vocabulary is further extended by including the hashtags corresponding to Food-101~\cite{bossard2014food} categories, resulting in a vocabulary of 1,170 unique hashtags. 

\textbf{Insta-1K dataset.} For each of the 1,170 unique hashtags,
at most 4250 images are downloaded from Instagram resulting in a total of $\sim3.7M$ images. Note that a single image could be retrieved and used for training several hashtags.
This dataset is referred to as \emph{Insta-1K} and used for training the auto-tagger. A subset of the dataset, called \emph{Insta-101}, consists of images belonging to the hashtags associated with the Food-101 categories. Assignment of hashtags to each of the Food-101 categories is performed manually. This dataset is used for performance comparisons on Food-101 categories. 


\textbf{County health statistics.} To see if signals derived from this online data are linked to patterns in the ``real world'', we obtained county-level health statistics for the year 2016 from the County Health Rankings and Roadmaps website~\footnote{\url{http://www.countyhealthrankings.org/rankings/data}}. This dataset includes statistics on various health measures that range from premature death and low birth weight to adult smoking, obesity and diabetes rates. From these statistics we decided to focus on the following nine health indicators: 
Smokers (in \%, lower is better), 
Adult Obesity (in \%, lower is better)
Food Environment Index (from 1=worst to 10=best, higher is better), 
Physically Inactive (in \%, lower is better), 
Excessive Drinking (in \%, lower is better), 
Alcohol-impaired Driving Deaths (in \%, lower is better), 
Diabetes Prevalence (in \%, lower is better), 
Food Insecurity (in \%, lower is better), and 
Limited Access to Healthy Food (in \%, lower is better).




\textbf{Food perception gap dataset.} We sampled images from the initial collection of $\sim4M$ Instagram posts associated with the US counties. Our county dictionary has 2,937 fips codes whereas the County Health Statistics dataset has 3,141 fips codes. Therefore, we used the 2,846 counties that were common in both datasets. 91 counties in our county dictionary without corresponding health statistics were dropped. 
We then kept the 194 counties with at least 2,000 posts. Finally, we removed images without at least one human tag appearing in at least 20 out of the 194 counties. This was done to remove images whose users might have very particular tagging behavior, resulting in a dataset of $1.9M$ posts used for food perception gap analyses.



\section{Machine Tagging}\label{sec:computervision}

For training the food auto-tagger we utilized the state of the art deep convolutional architectures called \emph{deep residual networks}~\cite{he2015deep}. These architectures have a proven record of success on a variety of benchmarks~\cite{he2015deep}. The main advantage of the deep residual networks is their residual learning framework which enables easier training of much deeper architectures (i.e.\ with 50, 101, 152 layers). The layers in the residual networks are reformulated as learning residual functions with reference to the layer inputs, instead of learning unreferenced functions as utilized in~\cite{simonyan2014very,krizhevsky2012imagenet}. Consequently, residual networks can train substantially deeper models which often result in better performances. 

\begin{table}
\begin{center}
\caption{Cross-dataset performances of deep residual networks on Food-101 categories. Note that the model performs $80.9\%$ accuracy on Food-101, $1.9\%$ higher than the previously reported state of the art~\cite{Myers:2015:ITA:2919332.2919978}.}
\vspace{1em}
\def\arraystretch{1.1}
\setlength{\tabcolsep}{.3em}
\begin{tabular}{l c|c|c|c}
 \parbox[t]{4mm}{\multirow{3}{*}{\rotatebox[origin=l]{90}{ Trained on}}}& & \multicolumn{2}{c}{Tested on} & \\
 & \multicolumn{1}{c|}{} & \multicolumn{1}{c|}{\bf Food-101} & \multicolumn{1}{c|}{\bf Insta-101} & \multicolumn{1}{c}{\bf mean} \\
\cline{2-5}
 & { \bf Food-101} & 80.9 & 40.3 & 60.6\\
 & { \bf Insta-101} & 74.5 & 51.5 & 63.0\\
\end{tabular}
\end{center}
\label{tab:food_insta_101}
\vspace{-1em}
\end{table}

We particularly train the deep residual network with 50 layers obtained from~\cite{he2015deep}. For benchmarking analysis, the model is first trained on Food-101 and Insta-101 datasets. 
As it is also mentioned in \cite{Myers:2015:ITA:2919332.2919978}, with 750 training and 250 test samples per category, Food-101 is the largest publicly available dataset of food images. On the other hand, our Insta-101 dataset has 4,000 training and 250 test samples per category collected from Instagram, though they are not manually cleaned. The task is to classify images into one of the existing 101 food categories. 
In the training procedure, the final 1000-way softmax in the deep residual model is replaced with a 101-way softmax, and the model is fine-tuned on the Insta-101 and Food-101 datasets individually. Training the deep residual model on Food-101 dataset, resulted in $\sim2\%$ improvement over the previously reported state of the art \cite{Myers:2015:ITA:2919332.2919978}. This illustrates that our auto-tagging architecture is highly competitive. The accuracies of the models are reported in Table~\ref{tab:food_insta_101} for comparison. 

\begin{figure}[h]
  \includegraphics[width=\columnwidth]{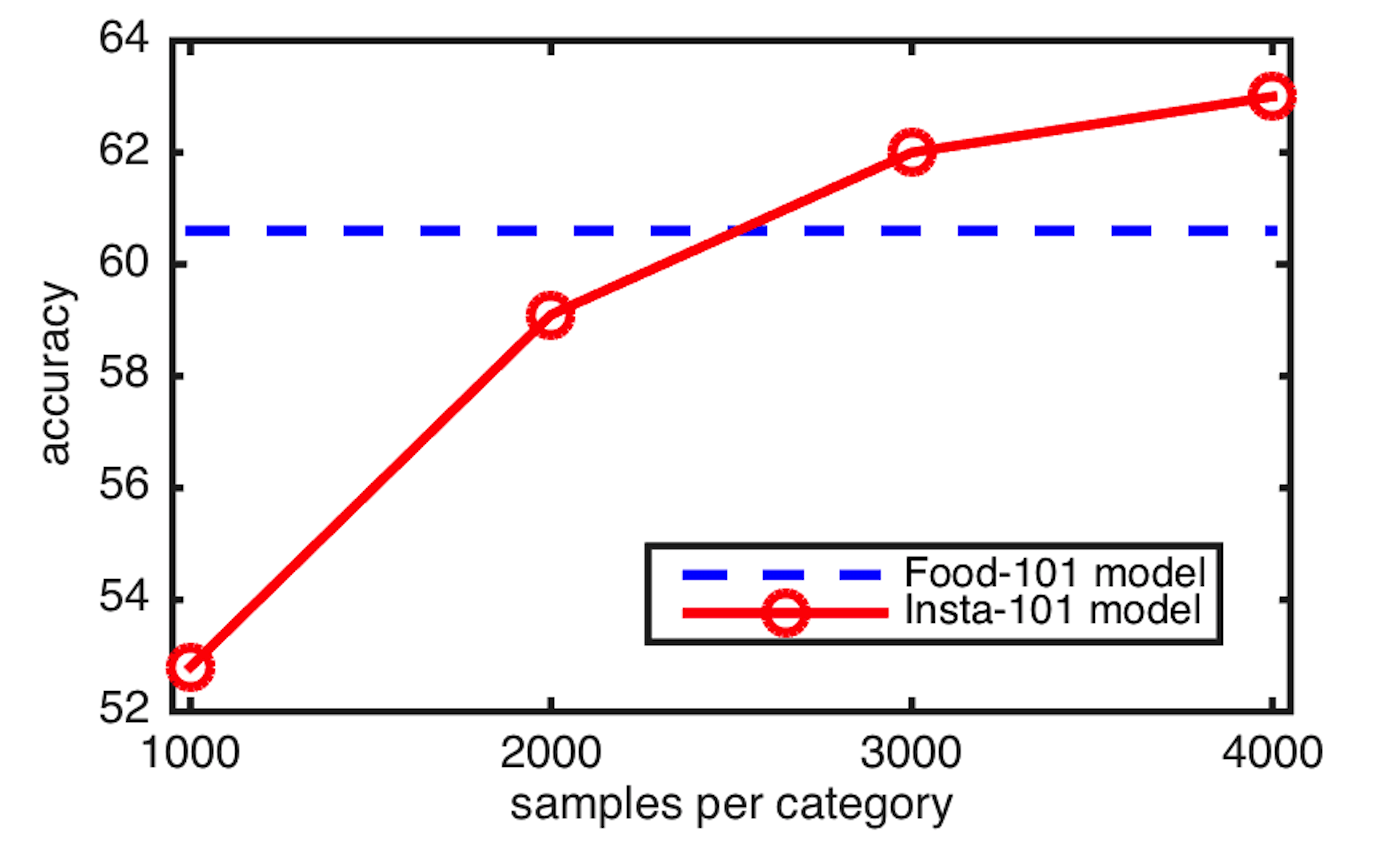}
  \caption{Mean cross-dataset performance of Insta-101 model trained with increasing number of training samples per category. Note that around 2500 samples per category Insta-101 model reaches the performance of Food-101 model.}
\label{fig:performance}
\end{figure}

The model trained with Insta-101 dataset performs remarkably well on Food-101 test set with an accuracy of $74.5\%$. Even though it is trained on the noisy social media images, on average our Insta-101-based classifier performs $\sim2.5\%$ better than the Food-101-based model, probably due to the increase in training samples from 750 to 4,000, which comes for free through Instagram query search.
We also report the mean cross-dataset performance of the Insta-101 model with increasing number of training samples in Figure~\ref{fig:performance}. Note that around 2500 samples per category the Insta-101 model reaches the performance of Food-101 model. The experiments on Food-101 categories suggest that, despite their noisy nature, Instagram images are valuable resources for training deep food classification models. 

The final auto-tagger is trained with the Insta-1K dataset, 1,170 categories with corresponding images collected from Instagram. Its accuracy is computed as $51.5\%$ on the held-out test set. For each image the top 30 hashtags identified by the auto-tagger are used for the perception gap analysis. 

\section{Methods}\label{sec:methods}

\subsection{Modeling Regional Variation in Health\\Statistics}\label{sec:methods:regional}
At a high level, our main analytical tool is simple correlation analysis of individual variables with ``ground truth'' county-level health statistics (see Section~\ref{sec:data}). This approach provides clues for hypotheses concerning causal links to explore further in separate studies.

In order to avoid spurious correlations, we perform 10-fold cross validation, leaving 19 (or 20) counties out and computing correlations using the rest of the 174 (or 173) counties at each fold. We then report average correlations and their standard errors in our results.

Also, when reporting significance values for $r$ correlation coefficients, we apply the Benjamini-Hochberge procedure~\cite{benjamini1995controlling} to guard against false positives. As an example, in Table~\ref{tab:objective} a significance level of $.05$ corresponds to a ``raw'' significance level of $.00328\pm.000075$.
%

We compute correlations for four different types of feature sets: (i) human tag usage probabilities, (ii) machine tag usage probabilities, (iii) perception gap weights, and (iv) conditional probabilities for the usage of \#healthy, \#delicious, and \#organic given machine tags. These will be described in the following.

\subsection{Quantifying the Perception Gap}\label{sec:methods:objectivegap}
They key contribution of this work is the analysis of the ``perception gap'' and how it relates to health. Abstractly, we define the perception gap as the difference between how a machine and a human annotate a given image. Concretely, we compute it as follows.

First, we iterate over all images. Images that do not have at least one machine tag and at least one human tag in the same vocabulary space (comprising the set $T$ of 1,170 tags described in Section~\ref{sec:computervision}) are ignored.

This was done as it is hard to quantify the disagreement between two annotators when one annotator does not say anything or uses a different (hashtag) language.
For each valid image we normalize the weights $w_i$ for both the machine tags $T^m$ and the human tags $T^h$ to probabilities such that $\sum_{i\in T^h} w^h_i = 1 = \sum_{i \in T^m} w^m_i$. Values for $w^h_i$ where $i \in T \setminus T^h$ or for $w^m_i$ where $i \in T\setminus T^m$ are set to 0. For $i \in T$ the gap value is then defined as $g_i = w^m_i - w^h_i$.
These values are then first averaged across all images for a (county,user) pair. We first aggregate at the user level, within a given county, to avoid that a single user with a particular hashtag usage pattern skews our analysis. Next, the user level values are further aggregated by averaging them into a single feature vector for each county. Note that each aggregated value $g_i$ will be between 0 and 1 and that $
\sum_{i\in T}|g_i|$ is a measure of the absolute overall labeling differences between humans and the machine in a given county. This difference is upper bounded by 2.

To obtain the human-only or machine-only distributions, we run the same filtering pipeline, simply setting $w^m_i=0$ (for human-only) or $w^h_i=0$ (for machine-only). Figure~\ref{fig:gap_ex1} gives an illustration by example of the qualitative aspects our perception gap can pick up. The image shown, which is a real example, is tagged by the user as \#foodie, \#hungry, \#yummy, \#burger, and by the machine as \#burger, \#chicken, \#fries, \#chips, \#ketchup, \#milkshake. After ignoring the user tags that are not included in our machine-tag dictionary, we compute the gap value $g_{\mbox{ex1}}$ for this particular image as

\[ \begin{cases} 
      -5/6 & \mbox{\#burger} \\
      1/6 & \mbox{\#chicken, \#fries, \#chips, \#ketchup, \#milkshake}\\
      0 & \mbox{all other hashtags} \\
   \end{cases}
\]\\


\subsection{Variation in Subjective Labels}\label{sec:methods:subjectivegap}
In the above, we computed a perceptual difference for ``what the food objectively is'' or for ``what is worth naming'', all related to \emph{objective} names of items in the picture. Here, we describe a methodology to compute a similar difference for \emph{subjective} labels.

As our machine annotations (see Section~\ref{sec:computervision}) deliberately exclude subjective tags, we can no longer use the previous approach of looking at human-vs.-machine usage differences within a common vocabulary space. Instead, we define a set of subjective labels of interest $l^h$ containing labels such as $j=$\#healthy, and then for each machine tag $i\in T^m$ compute the probability $P(j | i)$.

Concretely, we first iterate over all images who passed the filters for the previous ``objective gap'' analysis. For these images, we compute the aforementioned conditional probability probability at the image level where it is either 1 (if machine tag $i$ is present) or 0 (if it is not). Note that values for tags ${i'}\in T^m$ not present in the image are not considered. We then aggregate these values within each (county, user) pair to obtain probabilities for a given user in a given county to have used label $j$ given one of his images was auto-tag as $i$. We then further combine these probabilities at the county level by aggregating across users. If a tag ${i'}\in T$ was never present on a single image in the county then the corresponding value $P(j | {i'})$ is not defined. To address this, we impute such missing values by averaging the conditional expectation computed across all the counties with no missing values.
For our analysis we used the human labels \#healthy, \#delicious, and \#organic, comprising both health, taste and origin judgment.

\section{Results}\label{sec:results}
Table~\ref{tab:objective} shows the top five tags in terms of the ``boost'' they receive in correlation $r_{g_i}$ when using the gap values $g_i$ compared to the correlations $r_{w^m_i}$ and $r_{w^h_i}$ for features $w^m_i$ and $w^h_i$ respectively. Concretely, the boost is defined as $|r_{g_i}| - \max(|r_{w^m_i}|,|r_{w^h_i}|)$. 

\begin{table*}
\centering
\caption{For each of the nine health measures, top five tags with the highest correlation boost of the perception gap over the maximum of the correlations of machine-only and human-only tags. Values in parentheses are the mean and standard error of r correlation values across the 194 counties after 10-fold cross validation.}
\vspace{0.4em}
	\resizebox{\textwidth}{!}{%
	\begin{tabular}{|l|l|l|l|l|l|}
		\toprule
        Health metric & Top 1 & Top 2 & Top 3 & Top 4 & Top 5 \\ \midrule
AlcDrivDeath & chicagopizza ($-.24\pm.008$) & sugarcane ($.22\pm.010$) & brisket ($-.27\pm.006$) & redpepper ($.20\pm.009$) & mojito ($.22\pm.008$) \\
DiabetesPrev & crabmeat ($-.27\pm.007$) & burritos ($.29\pm.005$) & sushiroll ($-.31\pm.008$) & carpaccio ($-.23\pm.006$) & surfandturf ($.24\pm.009$) \\
ExcessDrink & beefribs ($-.24\pm.009$) & horseradish ($.25\pm.007$) & strawberries ($.24\pm.010$) & greeksalad ($.24\pm.010$) & flautas ($-.21\pm.007$) \\
FoodEnvInd & instabeer ($-.32\pm.010$) & heineken ($-.31\pm.006$) & cornedbeefhash ($-.26\pm.005$) & sushiroll ($.30\pm.009$) & jerkpork ($-.29\pm.007$) \\
FoodInsecure & instabeer ($.29\pm.008$) & beerstagram ($.25\pm.010$) & skirtsteak ($.27\pm.009$) & sushiroll ($-.29\pm.010$) & heineken ($.24\pm.007$) \\
LimitedAccess & breadpudding ($-.39\pm.008$) & horchata ($.46\pm.005$) & heineken ($.32\pm.009$) & quail ($-.31\pm.009$) & beersnob ($.28\pm.007$) \\
Obese & chickenkatsu ($.31\pm.007$) & clubsandwich ($-.33\pm.004$) & cobbsalad ($-.25\pm.005$) & moscato ($-.22\pm.005$) & koreanfriedchicken ($.22\pm.007$) \\
PhysInactv & prawns ($.34\pm.008$) & fishnchips ($.29\pm.009$) & burritos ($.28\pm.007$) & boilingcrab ($.30\pm.006$) & fishandchips ($.32\pm.007$) \\
Smokers & crabmeat ($-.30\pm.007$) & breadbowl ($.24\pm.006$) & umamiburger ($.26\pm.008$) & pupusas ($.24\pm.012$) & horchata ($.24\pm.010$) \\
        \bottomrule
	\end{tabular}
	}
	\label{tab:objective}
\end{table*}

As an example on how to read Table~\ref{tab:objective}, the entry ``chickenkatsu ($.31\pm.007$)'' as the Top 1 in the Obese row means that the counties where the machine is more likely than the human to use the tag \#chickenkatsu tend to have higher obesity rates. Furthermore, this correlation is significant at $p=.05$, even after applying the Benjamini-Hochberge procedure. The fact that values are ranked by their boost in correlation further means that the correlation of $.31\pm.007$ is not solely due to regional variation in what the machine tags as \#chickenkatsu. In this particular case, the machine-only correlation is $r_{w^m_i}=.19\pm.008$ and the human-only correlation is $r_{w^h_i}=.18\pm.007$.

Whereas Table~\ref{tab:objective} shows the results for the perception gap on objective tags (see Section~\ref{sec:methods:objectivegap}), Table~\ref{tab:subjective} shows results for the subjective gap (see Section~\ref{sec:methods:subjectivegap}). For this, tags from the space of the 1,170 machine tags are ranked according to the boost in correlation that the conditional probability of a human using, say, \#healthy achieves, compared to the correlation for the unconditional probability of a human using \#healthy.

\begin{table*}
\centering
\caption{For each of the nine health metrics and each of the three subjective tags $j\in  \{\mbox{healthy}, \mbox{delicious}, \mbox{organic}\}$ we show (up to) the top five tags $i$ in terms of correlation boost of using $P(j | i)$ over simply $P(j)$. Here $i$ is one of the 1,170 tags assigned by the machine. Only correlations significant at p=.05 (after applying the Benjamini-Hochberge procedure to guard against false positives) are shown. Values in parentheses are the mean and standard error of r correlation values across the 194 counties after 10-fold cross validation.}
\vspace{0.4em}
\begin{tabular}{lccc}
\toprule
 Health Metric & Healthy & Delicious & Organic \\
\midrule
\multirow{5}{*}{AlcDrivDeath} & cinnamonrolls ($-.24\pm.003$) & capresesalad ($-.30\pm.009$) & crepes ($-.27\pm.005$) \\
 & maple ($-.23\pm.005$) & breadsticks ($-.28\pm.007$) & popcorn ($-.25\pm.007$) \\
 & crawfish ($-.23\pm.008$) & beefcarpaccio ($-.28\pm.010$) & poachedegg ($-.24\pm.011$) \\
 & vanillabean ($-.23\pm.006$) & baguette ($-.27\pm.010$) & crawfish ($-.25\pm.009$) \\
 & coconutoil ($-.21\pm.008$) & redsauce ($-.27\pm.007$) & beefstew ($-.23\pm.003$) \\
\midrule
\multirow{5}{*}{DiabetesPrev} & smoothies ($-.30\pm.009$) & saki ($-.31\pm.007$) & jambalaya ($-.31\pm.006$) \\
 & thyme ($-.29\pm.005$) & crabmeat ($-.28\pm.006$) & carnitas ($-.31\pm.005$) \\
 & cilantro ($-.28\pm.008$) & burritos ($-.27\pm.009$) & chocolate ($-.31\pm.006$) \\
 & gyro ($-.27\pm.010$) & chimichanga ($-.26\pm.007$) & vinaigrette ($-.30\pm.004$) \\
 & bananas ($-.27\pm.009$) & octopus ($-.26\pm.006$) &  \\
\midrule
\multirow{5}{*}{ExcessDrink} & newyorkpizza ($.24\pm.006$) & barbecue ($-.23\pm.008$) & koreanfriedchicken ($.27\pm.008$) \\
 & bacardibuckets ($-.22\pm.015$) & burrito ($.22\pm.006$) & burritobowl ($.25\pm.005$) \\
 & cilantro ($.22\pm.009$) & winepairing ($.21\pm.009$) & appetizers ($.23\pm.005$) \\
 & dessert ($-.23\pm.005$) & beefcarpaccio ($-.19\pm.011$) & whiterice ($.23\pm.004$) \\
 & frenchpastry ($.20\pm.011$) & kalesalad ($.21\pm.010$) & cobbsalad ($.23\pm.007$) \\ 
\midrule
\multirow{5}{*}{FoodEnvInd} &  &  & carnitas ($.35\pm.006$) \\
 &  &  & seasalt ($.34\pm.006$) \\
 &  &  & coconutwater ($.32\pm.006$) \\
 &  &  & greendrink ($.32\pm.005$) \\
 &  &  & greenchile ($.31\pm.003$) \\
\midrule
\multirow{5}{*}{FoodInsecure} &  &  & carnitas ($-.35\pm.006$) \\
 &  &  & greendrink ($-.33\pm.005$) \\
 &  &  & seasalt ($-.33\pm.007$) \\
 &  &  & greenchile ($-.32\pm.003$) \\
 &  &  & kalesalad ($-.32\pm.008$) \\
\midrule
\multirow{5}{*}{LimitedAccess} & chorizo ($-.24\pm.008$) & takoyaki ($-.38\pm.004$) & beans ($-.22\pm.004$) \\
 & basil ($.22\pm.009$) & gin ($-.37\pm.005$) & southerncomfort ($-.22\pm.005$) \\
 & calzone ($-.21\pm.007$) &  & grits ($-.20\pm.006$) \\
 & catfish ($-.21\pm.012$) &  & grapefruit ($.19\pm.006$) \\
 & muffin ($.20\pm.017$) &  & onionrings ($-.19\pm.008$) \\
\midrule
\multirow{5}{*}{Obese} & sweettea ($-.35\pm.004$) & saki ($-.37\pm.007$) &  \\
 & chickpea ($-.34\pm.006$) & taquitos ($-.35\pm.006$) &  \\
 &  & smokedsalmon ($-.33\pm.006$) &  \\
 &  & chimichanga ($-.33\pm.006$) &  \\
 &  & fishtaco ($-.33\pm.007$) &  \\
\midrule
\multirow{5}{*}{PhysInactv} & chickpea ($-.20\pm.011$) & saki ($-.21\pm.009$) &  \\
 & maple ($-.22\pm.006$) & babybackribs ($.20\pm.008$) &  \\
 & cashews ($-.18\pm.010$) & wine ($-.20\pm.010$) &  \\
 &  & pitayabowls ($-.19\pm.006$) &  \\
 &  & bbqsauce ($.19\pm.005$) &  \\
\midrule
\multirow{5}{*}{Smokers} & goatcheese ($-.33\pm.010$) & saki ($-.41\pm.005$) &  \\
 & chickpea ($-.29\pm.008$) & soupoftheday ($-.37\pm.003$) &  \\
 & banana ($-.23\pm.008$) & milkshakes ($-.37\pm.006$) &  \\
 & roastbeef ($-.21\pm.011$) & smokedsalmon ($-.34\pm.007$) &  \\
 & cupcake ($-.21\pm.009$) &  &  \\
\bottomrule
\end{tabular}
\label{tab:subjective}
\end{table*}

As an example for how to read Table~\ref{tab:subjective}, the entry ``smoothies ($-.30\pm.009$)'' in the column for \#healthy and the row for Diabetes Prevalence means that counties with a higher conditional probability of P(human says \#healthy | machine says \#smoothies) tend to be counties with higher levels of diabetes prevalence. As we rank by the boost in correlation over the probability for simply P(human says \#healthy), in this case $-.30\pm.009$ vs.\ $-.27\pm.011$, this correlation is not fully explained by variation in \#healthy alone. As before, only correlations significant at p=.05, after the Benjamini-Hochberge procedure, are included in the table.

\section{Discussion and Limitations}\label{sec:discussion}
By looking at the ``healthy'' and ``organic'' columns in Table~\ref{tab:subjective} we see that, with the exception of Excessive Drinking, all health statistics indicate correlations in the good direction for all the examples shown. At a high level this seems to indicate that when humans deliberately call one particular food \#healthy or \#organic, rather than using these tags indiscriminately, this indicates a county with generally better health statistics. However, the pattern for what exactly this particular food has to be is far more mixed.


Similarly for the perception gap analysis in Table~\ref{tab:objective}, we observe correlations in the (reasonably) good direction in general. For instance, ``the machine says \#chickenkatsu (or \#koreanfriedchicken for that matter) but the human does not'' is a sign of a high obesity region while ``the machine says \#clubsandwich (or \#cobbsalad for that matter) but the human does not'' is a sign of a low obesity region. Similarly for diabetes prevalence, \#burritos shows positive correlation whereas \#crabmeat and \#sushiroll show negative correlation. However, in many other cases it is admittedly harder to interpret the results. For example, whereas ``the machine says \#sugarcane but the human does not'' is a sign of high alcohol-impaired driving deaths, for \#chicagopizza the trend is ``the human says \#chicagopizza but the machine does not.'' Likewise, the link between physical inactivity and hashtags such as \#prawns and \#fishnchips is not apparent.

It is  worth clarifying how our work differs fundamentally from analyzing the co-occurrence of hashtags. For example, we could hypothetically have studied how \#burger and \#salad are used together and whether their co-occurrence propensity was linked to health statistics. For the sake of argument, let us assume that we would have found that a positive association was linked to counties with lower obesity rates. However, we would then not have been able to tell if (i) healthier regions have more people consuming burgers with a salad on the side, or if (ii) in healthier regions people are simply more likely to label a lone lettuce leaf as \#salad. If the purpose of the analysis was to model regional variation in health statistics then this distinction might be irrelevant. But if the goal was to detect relationships between the perception of food and health statistics -- as is the case in our work -- then this distinction is crucial.

Note that, at the moment, we deliberately trained the machine tagger only on \emph{objective} tags related to food, e.g., the name of a dish. Training a machine for more \emph{subjective} tags, such as \#delicious or \#healthy, would have made it impossible to separate the dimension of ``what is it'' from ``how does a human perceive it''. However, it might be promising to train a machine on aspects related to the \emph{presentation} of food. As discussed in Section~\ref{sec:related}, how the food is presented to the consumer has important implications on how much of it will be consumed. When the food is presented and arranged by the consumer themselves, e.g., in the setting of a home-cooked meal, this could still provide a signal on whether the food is ``celebrated'' or not in a gourmet vs.\ gourmand kind of fashion.

One potential limitation of our work is language dependency. As we cannot look into users' brains to study the perception at the level of neurons, we rely on how they \emph{self-annotate} their images. However, a Spanish-speaking person will likely use other annotations than an English-speaking person, which eventually affects our analyses. For example, \#chimichanga and \#taquitos show up in our analysis as indicators of low obesity rates in the column for Delicious and row for Obese in Table~\ref{tab:subjective} even though both of them are deep-fried dishes from Mexican, or Tex-Mex, cuisine. Similarly, there can be regional variations and the same food could have one name in a high obesity area and a different name in a low obesity area.

Another risk comes from the inherent noise of the machine annotation used. Though its performance is state-of-the-art (see Section~\ref{sec:computervision}), it is still far from perfect. In the extreme case, if the machine annotations were uncorrelated with the image content then the perception gap we are computing would, on average, simply be the distribution of the human tags. As such, the gap and the human features would be picking up the same signals.

To guard against the previous two points we never solely report results for the perception gap analysis but, always, compare it back to the results when using only human annotations or only machine annotations. Both Table~\ref{tab:objective} and Table~\ref{tab:subjective} are ranked by the \emph{boost} in correlation over using only human annotations or only machine annotations.

Though our current analysis focuses on \emph{image} analysis, it is worth contemplating what a similar analysis of \emph{text} would look like. At a high level, we try to separate ``what something contains'' from ``how it is described''. In the NLP domain, this roughly corresponds to differentiating between topic detection \cite{wang2012baselines} and writing style identification \cite{ASI:ASI20316}. As for images, these two are often entangled: if an author of a blog uses the term ``death'' is that because of (i) the topic they are discussing (e.g., a war story), or because of (ii) their writing style and mood (potentially indicating depression)? Clearly separating these two concepts, the what and the how, could potentially help with challenges such as identifying depression.

The current work uses image recognition exclusively to study how a certain food is labeled by a human in relation to what it objectively shows. However, a computer vision approach could be used for automating other aspects of analyzing food images. As an example, it could be promising to automatically analyze \emph{food plating}, i.e., the aesthetic arrangement of food in appealing images. Recent research studies have indicated that attractive food presentation enhances diners' liking of food flavor~\cite{michel2014taste, zellner2014tastes} as well as their eating behaviours and experiences~\cite{spence2014plating}. In addition, Zampollo et al.~\cite{zampollo2012looks} have demonstrated the diversity of food plating between cultures. Extending these ideas in mind, a computer vision approach could be applied to perform a study to that of Holmberg et al.~\cite{holmberg2016adolescents} to investigate food plating (and its potential correlation with food health) across cultures and age groups.

Both regional variation in perception gap and in food plating behavior could conceptually be used for public health monitoring by training models similar to what was done by Garimella et al.~\cite{garimellaetal16chi}. We briefly experimented with this using our $g_i$ gap features. However, these secondary signals, i.e., how something is perceived differently by humans and machines, did not add predictive performance over the primary signals, i.e., how something is labeled by humans or machines alone. We see the real value of our approach less in ``now-casting'' of public health statistics and more in analyzing the psychology of food consumption and food sharing.

Finally, computer vision could help to obtain health labels not only at the county level but at the \emph{individual} level. Concretely, Wen and Guo have proposed a method to infer a person's BMI from a clean, head-on passport style photo~\cite{wenguo13ivc}. Though this particular method is unlikely to deal with the messiness of social media profile images, exploratory work has shown that inferring weight category labels from social media profile images seems feasible~\cite{webermejova16dh}. We are currently working on these aspects of a more holistic food image scene understanding. 

\section{Conclusions}

In this work, we define the ``perception gap'' as the misalignment or the difference in probability distributions of how a human annotates images vs.\ how a machine annotates them. To the best of our knowledge, this is the first time that this type of perception gap has been studied. By using county-level statistics we show that there are systematic patterns in how this gap relates to health outcomes.

In particular we find evidence for the fact that conscious food choices seem to be associated with regions of better health outcomes. For example, labeling particular foods as \#healthy, rather than random images in a county, or beer with its brand name, rather than generic descriptions, correlates favorably with health statistics. Similarly, posting images of saki and emphasizing the \#delicious taste appears to be a positive indicator. Paraphrasing Shakespeare, a rose by any other name might smell as sweet, but labeling your food differently might be related to health.

As time goes on, we expect our methodology to further improve in performance due to (i) continuous improvement in image recognition and decrease in error rates, and due to (ii) the potential to use individual level health labels, instead of county level ones, also due to improvement in computer vision \cite{wenguo13ivc,webermejova16dh}.

%

%

\end{document}